\documentclass[pdftex,twocolumn,epjc3]{svjour3}   
\RequirePackage[T1]{fontenc}

\smartqed  

\RequirePackage{graphicx}
\RequirePackage{flushend}
\RequirePackage[numbers,sort&compress]{natbib}
\RequirePackage[colorlinks,citecolor=blue,urlcolor=blue,linkcolor=blue]{hyperref}
\RequirePackage{amsmath}
\RequirePackage{multirow}
\RequirePackage{xspace}
\RequirePackage{xcolor}
\RequirePackage{cancel}
\RequirePackage[caption=false]{subfig}
\RequirePackage{subfig}
\RequirePackage{blindtext, subfig}
\RequirePackage{amsmath}
\usepackage{amssymb}
\usepackage{dblfloatfix}% http://ctan.org/pkg/dblfloatfix
\usepackage{lipsum}% http://ctan.org/pkg/lipsum
\usepackage{ulem}
\RequirePackage{fix-cm}

\RequirePackage{comment}
\RequirePackage{hyperref}
\RequirePackage{slashed}
\RequirePackage[caption=false]{subfig}
\RequirePackage{tabu}
\usepackage{hyperref}
\usepackage{multicol}
\usepackage[super]{nth}
\usepackage{booktabs}
\usepackage{dutchcal} % \mathcal: also small letters. \mathbcal = bold ones.

\newcommand{\MGvATNLO}{{\tt {\sc MadGraph5}\_aMC@NLO}}

\begin{document}

\title{Axion-Like Particles at future $e^- p$ collider}

\author{Karabo Mosala\thanksref{e1,addr1} \and Pramod Sharma\thanksref{e2,addr1,addr2} \and Mukesh Kumar\thanksref{e3,addr1} \and Ashok Goyal\thanksref{e4,addr3}%etc.
}
\thankstext{e1}{e-mail: \textcolor{blue}{karabo.mosala@cern.ch}}
\thankstext{e2}{e-mail: \textcolor{blue}{pramodsharma.iiser@gmail.com}}
\thankstext{e3}{e-mail: \textcolor{blue}{mukesh.kumar@cern.ch}}
\thankstext{e4}{e-mail: \textcolor{blue}{agoyal45@yahoo.com}}

\institute{School of Physics and Institute for Collider Particle Physics, University of the Witwatersrand, Johannesburg, Wits 2050, South Africa. \label{addr1}
           \and
Indian Institute of Science Education and Research, Knowledge City, Sector 81, S. A. S. Nagar, Manauli PO 140306, Punjab, India. \label{addr2}
           \and
Department of Physics \& Astrophysics, University of Delhi, Delhi 110 007, India.\label{addr3}
}

\date{Received: date / Accepted: date}

\maketitle

\begin{abstract}
In this work, we explore the possibilities of producing Axion-Like Particles (ALPs) in a future $e^-p$ collider. Specifically, we focus on the proposed Large Hadron electron collider (LHeC), which can achieve a center-of-mass energy of $\sqrt{s} \approx 1.3$~TeV, enabling us to probe relatively high ALP masses with $m_a \lesssim 300$~GeV. The production of ALPs can occur through various channels, including $W^+W^-$, $\gamma\gamma$, $ZZ$, and $Z\gamma$-fusion within the collider environment. To investigate this, we conduct a comprehensive analysis that involves estimating the production cross section and constraining the limits on the associated couplings of ALPs, namely $g_{WW}$, $g_{\gamma\gamma}$, $g_{ZZ}$, and $g_{Z\gamma}$. To achieve this, we utilize a multiple-bin $\chi^2$ analysis on sensitive differential distributions. Through the analysis of these distributions, we determine upper bounds on the associated couplings within the mass range of 5~GeV $\leq m_a \leq$ 300~GeV. The obtained upper bounds are of the order of ${\cal O}(10^{-1})$ for $g_{\gamma\gamma}$ ($g_{WW}$, $g_{ZZ}$, $g_{Z\gamma}$) in $m_a \in$~[5, 200 (300)]~GeV considering an integrated luminosity of 1~ab$^{-1}$. Furthermore, we compare the results of our study with those obtained from other available experiments. We emphasize the limits obtained through our analysis and showcase the potential of the LHeC in probing the properties of ALPs.

\end{abstract}

\section{Introduction}
\label{sec:intro}
Axion-like particles (ALP) are Standard Model (SM) singlet pseudo-scalar Nambu-Goldstone bosons originally proposed to solve the strong $CP$ problem~\cite{Peccei:1977hh, Peccei:1977ur, Weinberg:1977ma, Wilczek:1977pj}. Their interaction with the SM particles arises from an explicit breaking of an approximate Global Peccei-Quinn $U(1)_{PQ}$ symmetry with couplings considered as free parameters (see for example~\cite{Alonso-Alvarez:2018irt, Bonilla:2021ufe}). Subsequently, ALPs made their appearance in beyond the SM (BSM) $viz.$ composite models~\cite{Kim:1984pt, Leder:1996py}, Grand Unification models~\cite{Feindt:1991rb, Georgi:1981pu, Rubakov:1997vp}, extra-dimension models~\cite{Dienes:1999gw, Legero_2003}, super- symmetric models~\cite{Lillard:2017cwx}, string theories~\cite{Svrcek:2006yi} etc. The axion mass and the new physics scale associated with the new physics varied over the vast range of ALP mass (sub eV to TeV) with the scale varying from electroweak to TeV and beyond. Light ALPs with masses less than eV to MeV range can modify the Cosmic Microwave Background (CMB), big Bang Nucleosynthesis (BBN), cooling and evolution of stars. Their coupling to SM particles within this mass range is severally constrained from astrophysical and cosmological observations~\cite{PhysRevD.104.035038, PhysRevD.104.103521, Green:2017ybv, Baumann:2016wac, Krnjaic:2019dzc, Raffelt:2006cw, Lee:2018lcj, Chang:2018rso}. Heavier ALPs in the MeV to TeV mass range though, unimportant from the astrophysical and cosmological considerations, have been the subject matter of recent studies in the context of particle physics and as dark matter portals connecting the dark-matter with the visible matter~\cite{Darme:2020sjf, Agrawal:2021dbo} and have been employed to explain leptonic $g-2$ anomaly with some success~\cite{Long:2015pua, BESIII:2020pxp}. The mass and interaction of these particles with the visible matter have been explored at the high energy colliders like LEP, Tevatron, Belle-II and at the CERN Large Hadron Collider (LHC) (see for example~\cite{dEnterria:2021ljz,Brivio:2017ije,Bauer:2017ris,Chala:2020wvs,Bauer:2018uxu,Bauer:2020jbp,Bauer:2021mvw,Biekotter:2023mpd,Alonso-Alvarez:2018irt}). Unlike these colliders, the future $e^{+}e^{-}$ colliders like ILC~\cite{Zhang:2021sio}, FCC-ee~\cite{TLEPDesignStudyWorkingGroup:2013myl, Abada:2019}, CEPC~\cite{Bozovic-Jelisavcic:2018zhc, CEPCStudyGroup:2018ghi, An:2018dwb}, the high energy muon collider~\cite{MuonCollider:2022nsa} and the electron-hadron collider (LHeC) using the LHC protons on the electron beam~\cite{LHeCStudyGroup:2012zhm, LHeCStudyGroup:2012wne} are designed to have high luminosity, and they provide cleaner experimental environment to go beyond the LHC’s precision ability and are eminently suited to determine the ALP properties. Constraints on a large range of ALP parameter space were established from LHC and the future $e^+ e^-$ collider experiments~\cite{Jaeckel:2012yz, Baldenegro:2018hng, Florez:2021zoo, Marciano:2016yhf, Inan:2020aal, Inan:2020kif, Zhang:2021sio, Buttazzo:2018qqp, Gavela:2019cmq} through the photon fusion production to obtain the sensitivities on the ALP $\gamma\gamma$ coupling for the ALP mass in the 1~GeV to $\sim600$~GeV range. The possibility of detecting ALP production through electro-weak massive vector-boson fusion (VBF) processes was recently investigated in the future muon collider for $m_a$ ${\cal O}$(TeV) and beyond to study the $WW$, $ZZ$, $Z\gamma$ and $\gamma \gamma$ coupling constraints~\cite{Han:2022mzp}.

In this work, we investigate the possibility of detecting ALPs production $via$ VBF processes at future Large Hadron-electron Collider (LHeC) $e^{-}p$ colliders, focusing on producing constraints on possible couplings parameters, $g_{\gamma \gamma}$, $g_{WW}$, $g_{Z\gamma}$ and $g_{ZZ}$~\cite{Yue:2023mew, Yue:2019gbh}. We base our study on LHeC environment, which employs the 7~TeV proton beam of the LHC and electrons from an Energy Recovery Linac (ERL) being developed for the LHeC. The choice of an ERL energy of $E_{e} = 60 (120)$ GeV with an available proton energy $E_{p} = 7$ TeV would provide a centre of mass energy of $\sqrt{s} \approx 1.3 (1.8)$~TeV at the LHeC~\cite{LHeCStudyGroup:2012zhm, Bruening:2013bga, LHeCStudyGroup:2012wne}.

This article is organised in following sections: in~\autoref{sec:model} model with effective Lagrangian and analysis framework is explained, a preliminary estimation of ALP production as a function of $m_a$, coupling(s) and LHeC energies are explored in~\autoref{sec:prod}, and results using different observable(s) are explained in~\autoref{sec:results}. The comparison(s) of ours findings with existing results are discussed in~\autoref{comp} and a summary with discussions are followed in~\autoref{sec:sum}.   

\section{Model and Framework}
\label{sec:model}
The interactions of ALPs with gauge bosons and SM fermions occurs via the dimension five operators, with their masses considered independently of their respective coupling strengths \cite{Brivio:2017ije}. Hence the effective interactions between the ALPs and the electroweak gauge bosons are represented by the effective Lagrangian~\cite{Bauer:2017ris,Bauer:2018uxu,Georgi:1986df}:
 \begin{align}
 \label{eq:1}
{\cal L}_{\rm eff} =&\, \frac{1}{2}(\partial{_{\mu} a})(\partial^{\mu} a) - \frac{1}{2} m_a^2 a^2 + g^2 C_{WW} \frac{a}{f_a} W^{A}_{\mu \nu} \tilde{W}^{\mu \nu A} \notag\\
&\,+ g^{\prime 2} C_{BB} \frac{a}{f_a} B_{\mu \nu} \tilde{B}^{\mu \nu},
\end{align}
where $X_{\mu\nu}$ represents the field strength tensor for the $SU(2)_L$ or $U(1)_Y$, $\tilde{X}^\mu{}^\nu = \frac{1}{2}\varepsilon^{\mu \nu \alpha \beta} X_{\alpha \beta}$ with $\varepsilon^{0123} = 1$ and $X \in \{B,W\}$. The ALP field and mass are represented by $a$ and $m_{a}$, respectively. After electroweak symmetry breaking we can write  the interactions between the ALP and the electroweak gauge bosons ($W^\pm$, $Z$, $\gamma$) in terms of  dimension-less couplings $g_{\gamma\gamma}$, $g_{WW}$, $g_{Z\gamma}$ and $g_{ZZ}$ respectively as:
\begin{align}
\label{eq:2}
{\cal L}_{\rm eff} \supset&\, e^{2}\frac{a}{f_a} g_{\gamma\gamma} F_{\mu\nu}\Tilde{F}^{\mu\nu} + \frac{2e^{2}}{c_w s_w} \frac{a}{f_{a}} g_{Z\gamma} F_{\mu\nu}\Tilde{Z}^{\mu\nu} \notag\\
&+ \frac{e^{2}}{c^2_w s^{2}_w} \frac{a}{f_a} g_{ZZ} Z_{\mu\nu}\Tilde{Z}^{\mu\nu} +\frac{e^{2}}{s^{2}_w} \frac{a}{f_a} g_{WW} W_{\mu\nu}\Tilde{W}^{\mu\nu}.
\end{align}
In terms of $C_{ij}$ ($i,j \equiv \gamma, Z, W$), the couplings $g_{ij}$ are given by
\begin{align}
\label{eq:3}
\left.\begin{array}{l}
  g_{\gamma \gamma} = C_{WW}+C_{BB}, \\ \\ g_{Z\gamma} = c_w^2 C_{WW}-s_w^2 C_{BB}, \\ \\
  g_{ZZ} = c_w^4 C_{WW} + s_w^4 C_{BB},  \\ \\ g_{WW} = C_{WW}.
\end{array}\right\},
%\left.\begin{array}{lc}
%  g_{\gamma \gamma} = C_{WW}+C_{BB},   & g_{Z\gamma} = c_w^2 C_{WW}-s_w^2 C_{BB}, \\
%  &\\
%  g_{ZZ} = c_w^4 C_{WW} + s_w^4 C_{BB},   & g_{WW} = C_{WW}.
%\end{array}\right\},
\end{align}
where $c_w$ and $s_w$ are the cosine and sine of the Weinberg mixing angle $\theta_w$, respectively. For all studies in this work the scale parameter is fixed to $f_a = 1$~TeV.
% In this work, for simplicity all couplings $g_{ij}$ are fixed at 1~TeV$^{-1}$ with the scale parameter is fixed to $f_a = 1$~TeV.
%
\begin{figure}[t]
\centering
{\includegraphics[width=.45\textwidth]{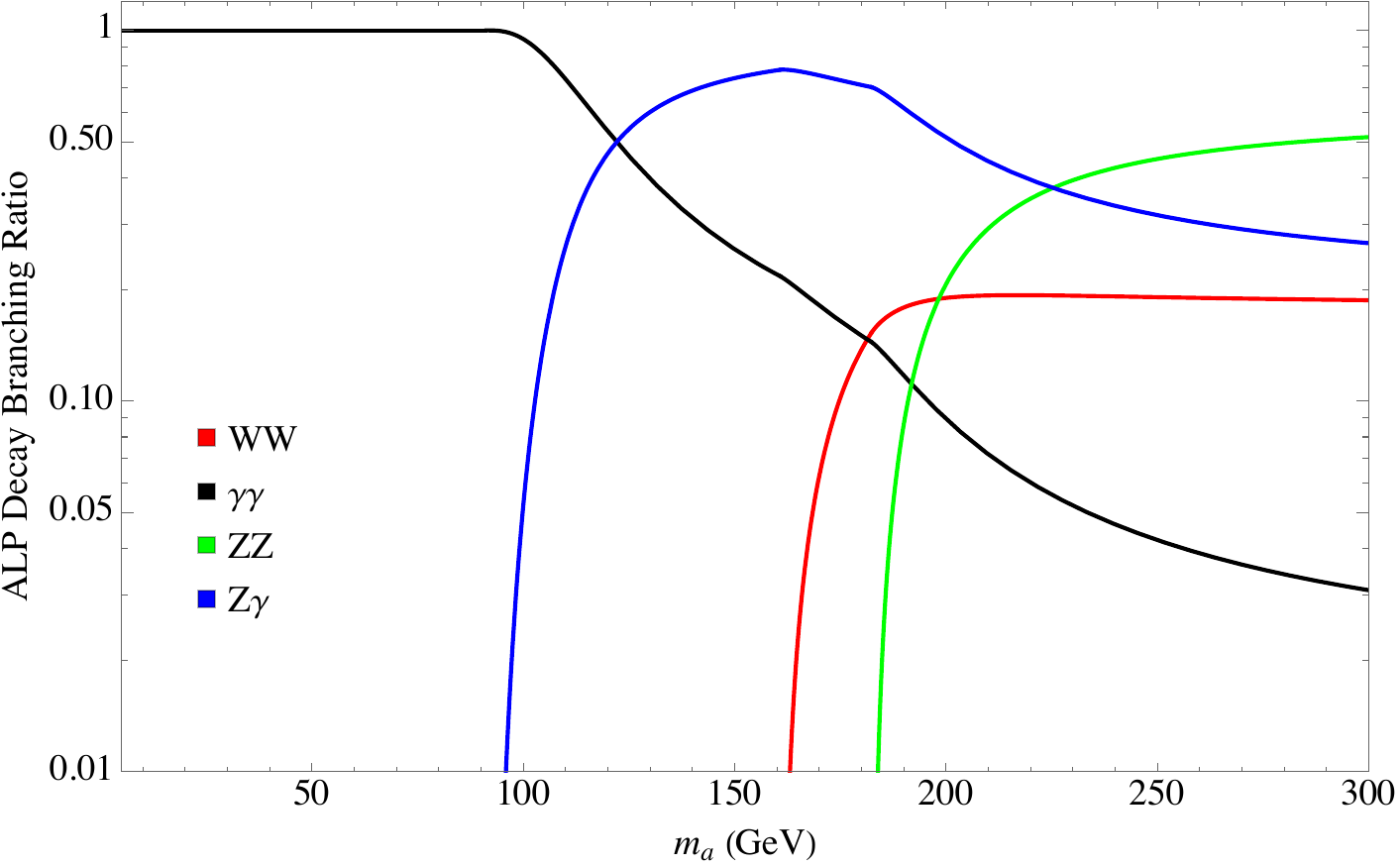}}
\caption{The branching ratios for the decay modes of a massive ALP, $a \to W^+W^-$, $\gamma\gamma$, $ZZ$ and $Z\gamma$ as a function of its mass $m_a$ by keeping the couplings $g_{ij} = 1$ and the scale parameter $f_a = 1$~TeV.}
\label{fig:br}
\end{figure}
\begin{figure*}[t]
\centering
\subfloat[\label{fig:WW}]{\includegraphics[width=.25\textwidth]{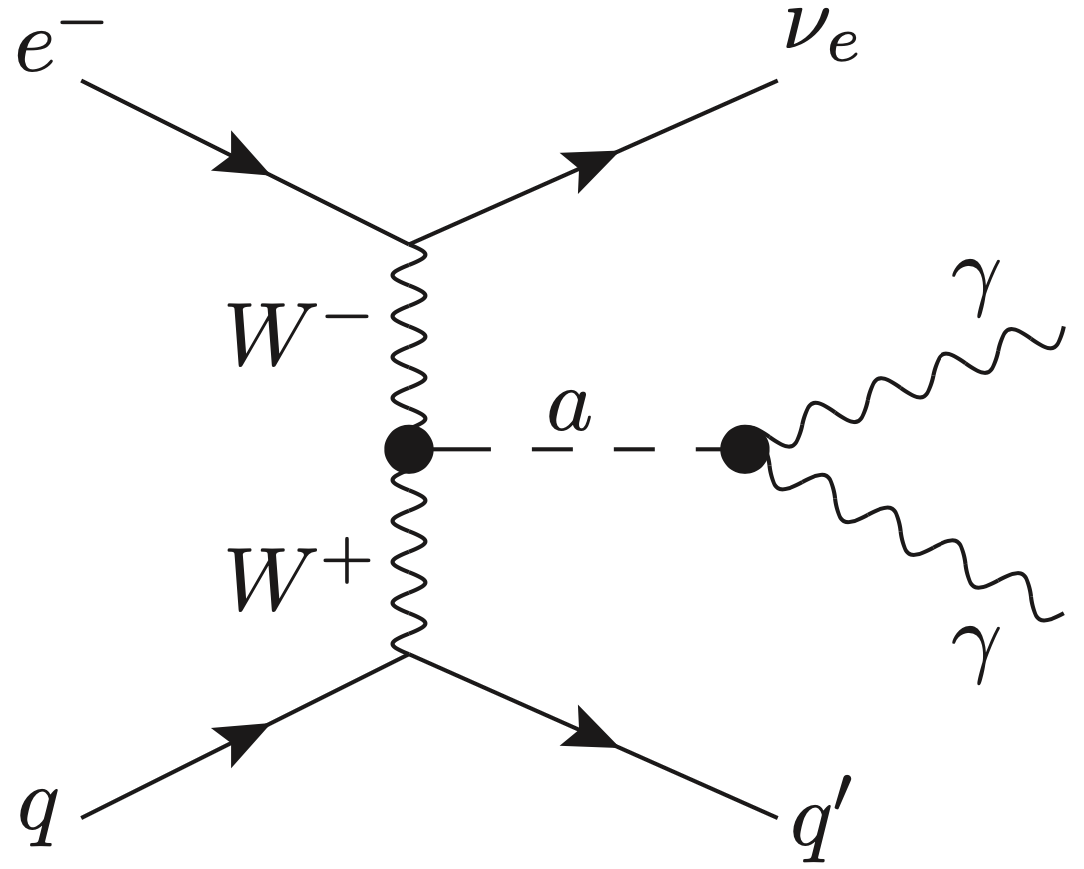}}
\subfloat[\label{fig:yy}]{\includegraphics[width=.25\textwidth]{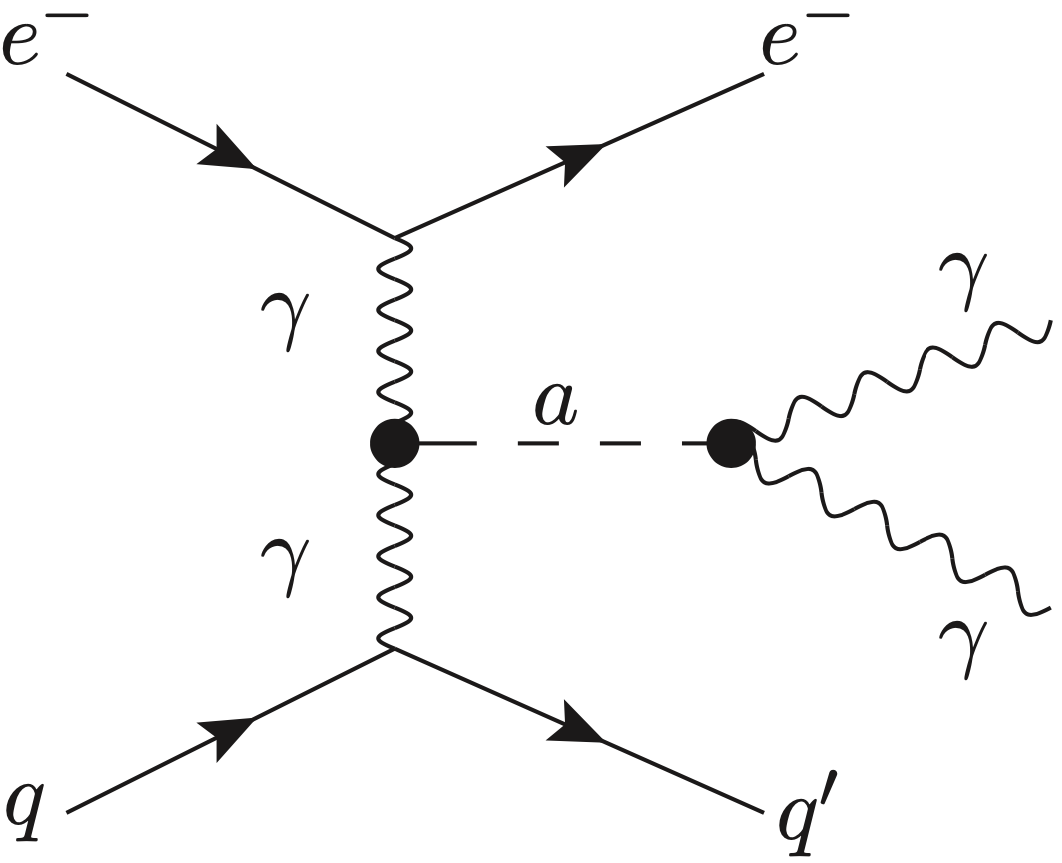}}
\subfloat[\label{fig:ZZ}]{\includegraphics[width=.25\textwidth]{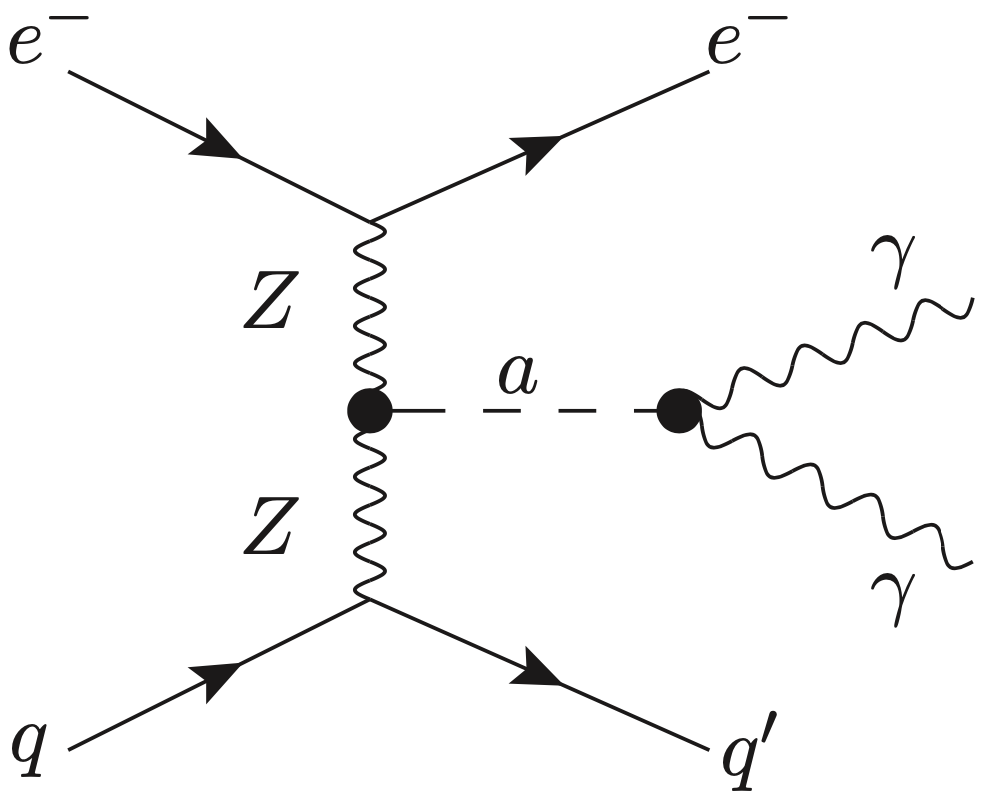}}
\subfloat[\label{fig:Zy}]{\includegraphics[width=.25\textwidth]{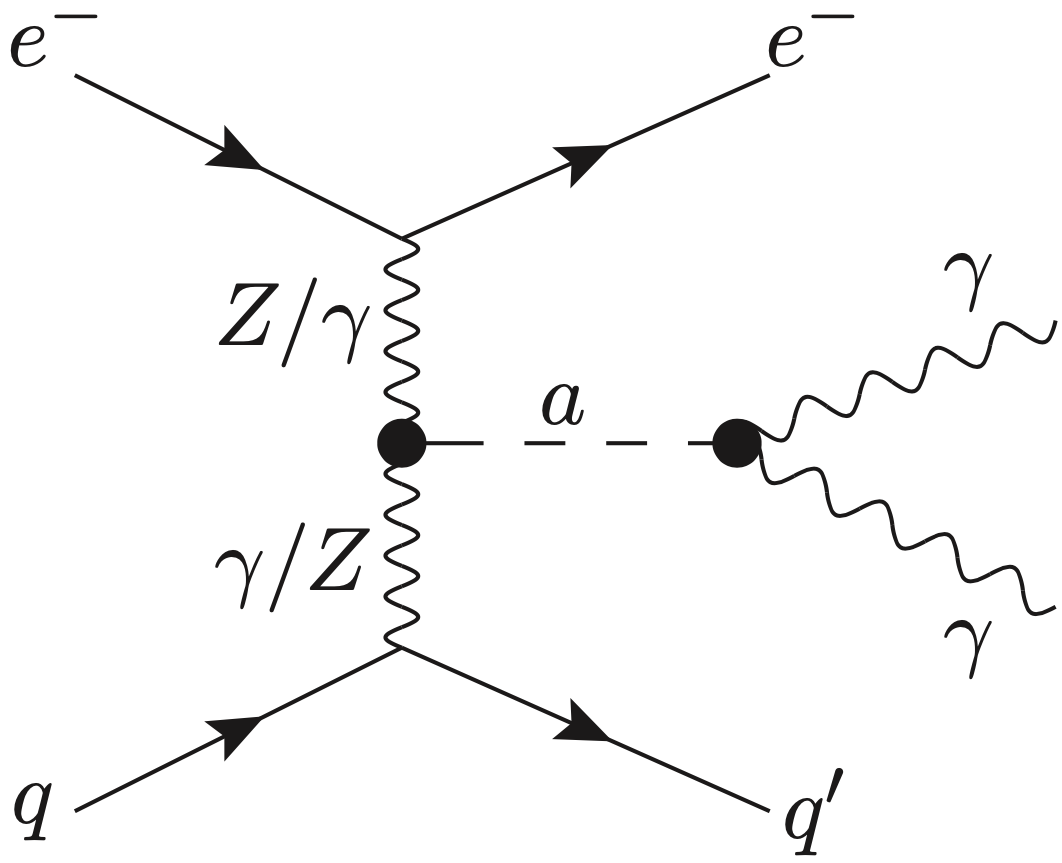}}
\caption{Leading order representative Feynman diagrams at matrix-element level for single ALP production in (a) {\tt CC} through $W^\pm$-fusion [{\tt{WW}}] and {\tt NC} through (b) $\gamma\gamma$, (c) $ZZ$ [{\tt{ZZ}}], and (d) $Z\gamma$-fusion [${\tt{Z}}\gamma$] processes in deep inelastic electron-proton collisions. A particular decay of $a\to \gamma\gamma$ is considered in this study. Here, $q, q^\prime$ $\equiv$ $u,\bar{u}$, $d,\bar{d}$, $c,\bar{c}$, $s,\bar{s}$, or $b,\bar{b}$.\label{fig:1}}
\end{figure*}

Using the interactions defined in~\autoref{eq:2}, the relevant decay widths of ALP are given by
\begin{align}
    &\Gamma(a \to W^+W^-) \equiv \Gamma_{WW} \notag\\
&\qquad = \frac{e^4}{8 \pi f_a^2 s_{w}^4}\left| g_{WW}\right|^2 m_a^3 \left(1-4\frac{m_W^2}{m_a^2} \right)^{\frac{3}{2}}, \label{eq:4}
\end{align}
\begin{align}
    &\Gamma(a \to \gamma \gamma) \equiv \Gamma_{\gamma\gamma} = \frac{e^4}{4 \pi f_a^2}\left| g_{\gamma\gamma}\right|^2 m_a^3,\label{eq:5}
\end{align}
\begin{align}    
    &\Gamma(a \to Z Z) \equiv \Gamma_{ZZ} \notag\\
&\qquad = \frac{e^4}{4 \pi f_a^2 c_{w}^4 s_{w}^4}\left| g_{ZZ}\right|^2 m_a^3 \left(1-4\frac{m_Z^2}{m_a^2} \right)^{\frac{3}{2}}, \label{eq:6}
\end{align}
\begin{align}
    &\Gamma(a \to Z\gamma) \equiv \Gamma_{Z\gamma} \notag\\
&\qquad = \frac{e^4}{2 \pi f_a^2 c_{w}^2 s_{w}^2}\left| g_{Z\gamma}\right|^2 m_a^3 \left(1-\frac{m_Z^2}{m_a^2} \right)^3, \label{eq:7}
\end{align}
where $m_W$ and $m_Z$ represent the masses of the $W^\pm$ and $Z$ bosons, respectively. As $\Gamma_{ij}$ is a function of corresponding coupling and masses of ALP, in this study we take variable decay width to find the limits of $g_{ij}$ as a function of $m_a$. In \autoref{fig:br}, the branching ratios for the decay modes $a \to W^+W^-$, $\gamma\gamma$, $ZZ$, and $Z\gamma$ are plotted as a function of the mass of the ALP, $m_a$, assuming $g_{ij} = 1$.

Further, we define following formula to find local significance and discovery limits for a given number of signal ($S$) and background ($B$) events at a particular luminosity $L$, considering the total statistical and systematic uncertainties $\delta_{s}$ as
\begin{align}
 N_{\rm SD} = \frac{S}{\sqrt{S + B + (\delta_{s}\cdot S)^2 +(\delta_{s}\cdot B)^2}},\label{eq:8}    
\end{align}
where in terms of corresponding cross section of signal $\sigma(g_{ij})$ and background $\sigma_{\rm SM}$, $S = \sigma(g_{ij})\cdot L$ and $B = \sigma_{\rm SM}\cdot L$, respectively. 

Also to constrain the ALP$-$gauge coupling $g_{ij}$, we use a $\chi^2$-analysis both at total cross-section and most sensitive differential-distribution level, where the $\chi^2$ definition is given by
\begin{align}
    \chi^2 = \sum_{k=1}^{n} \left(\frac{N_k(g_{ij}) - N_k^{\rm SM}}{\Delta N_k}\right)^2. \label{eq:9}
\end{align}
In this case, $N_k(g_{ij})$ represents number events for signal in $k^{th}$ bin of a distribution of total $n$ bins while $N_k^{\rm SM}$ is the corresponding background and $\Delta N_k$ is defined as:
\begin{align}
   \Delta N_k = \sqrt{N_k^{\rm SM}\left(1+\delta_{s}^2 N_k^{\rm SM}\right)}. \label{eq:10}
\end{align}
For our results we consider $\delta_{s} = 5\%$ for a given luminosity $L$, and $L = 1$~ab$^{-1}$. 

\section{ALP production in $e^-p$ collider}
\label{sec:prod}

As mentioned in~\autoref{sec:intro}, we are interested to probe ALP-gauge couplings by direct production of ALP through VBF processes in $e^-p$ collider. In such an environment, using the interactions defined in~\autoref{sec:model} the direct production of ALP can occur in charged-current ({\tt CC}) mode through $W$-boson fusion as shown in~\autoref{fig:WW} [{\tt{WW}}], and in neutral-current ({\tt NC}) mode through $\gamma\gamma$~(\autoref{fig:yy}), $ZZ$~(\autoref{fig:ZZ} [{\tt{ZZ}}]) and $Z\gamma$-fusion~(\autoref{fig:Zy} [{\tt{Z}}$\gamma$]), where in particular we have considered the decay of ALP, $a\to \gamma\gamma$ (so we keep $g_{\gamma\gamma} = 1$ in all channels), for a given $m_a$. For all results, the branching ratio of ALP decay to di-photon ${\cal B}_{a\to\gamma\gamma}$ is taken as function of $m_a$, considering two cases: Case (I), where the corresponding channel's coupling is set to 1 and others to 0; and Case (II), where all couplings $g_{ij}$ are uniformly set to 1 as depicted in~\autoref{fig:br}. Here, we also note that the ${\tt Z}\gamma$-channel cannot be separated from the $\gamma\gamma$-channel and hence $g_{\gamma\gamma}\ne 0$; though  for Case (I) we can choose $g_{ZZ} = 0$.\footnote{Important to mention: for $m_a > m_Z$, ${\cal B}_{a\to \gamma\gamma} < 1$ as $a\to Z\gamma$ channel opens up (\autoref{fig:br}); and deviations will become apparent in any observable for Case (I) {\it{vs}} Case (II).} Therefore, the notation $Z\gamma$ will refer to the effect of considering the channels shown in~\autoref{fig:yy},~\autoref{fig:Zy} (and \autoref{fig:ZZ} in Case (II)), and their interference.

\begin{figure*}[t]
\centering
\subfloat[\label{xsec-ma}]{\includegraphics[width=.50\textwidth]{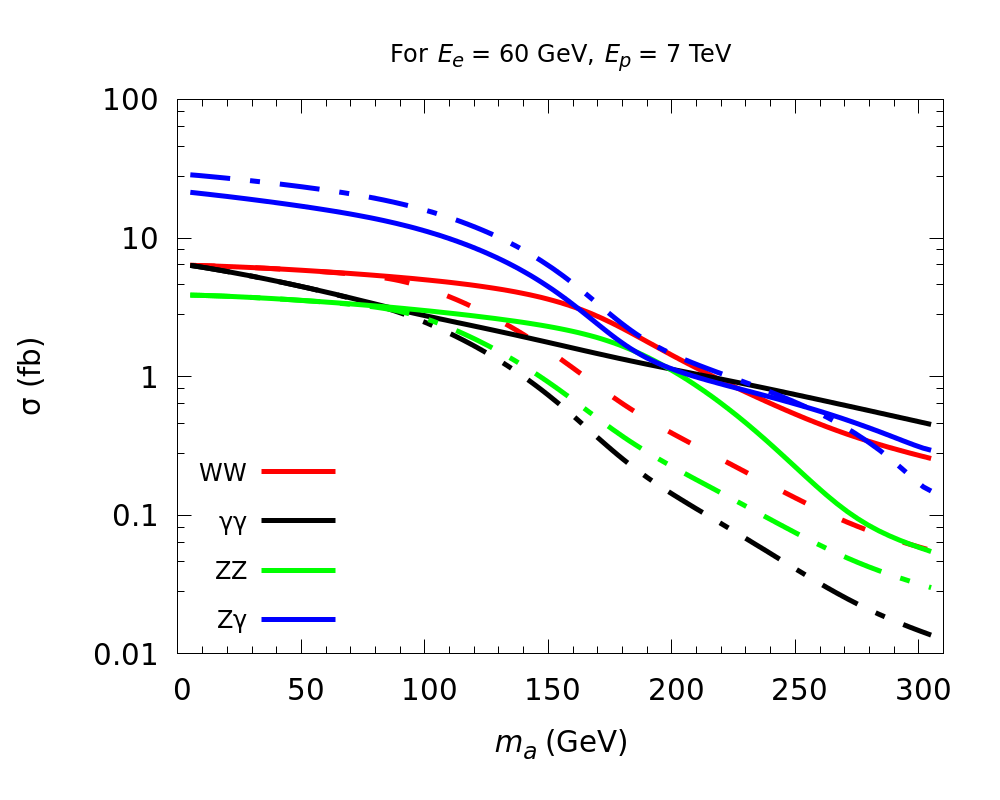}}
\subfloat[\label{xsec-ee}]{\includegraphics[width=.50\textwidth]{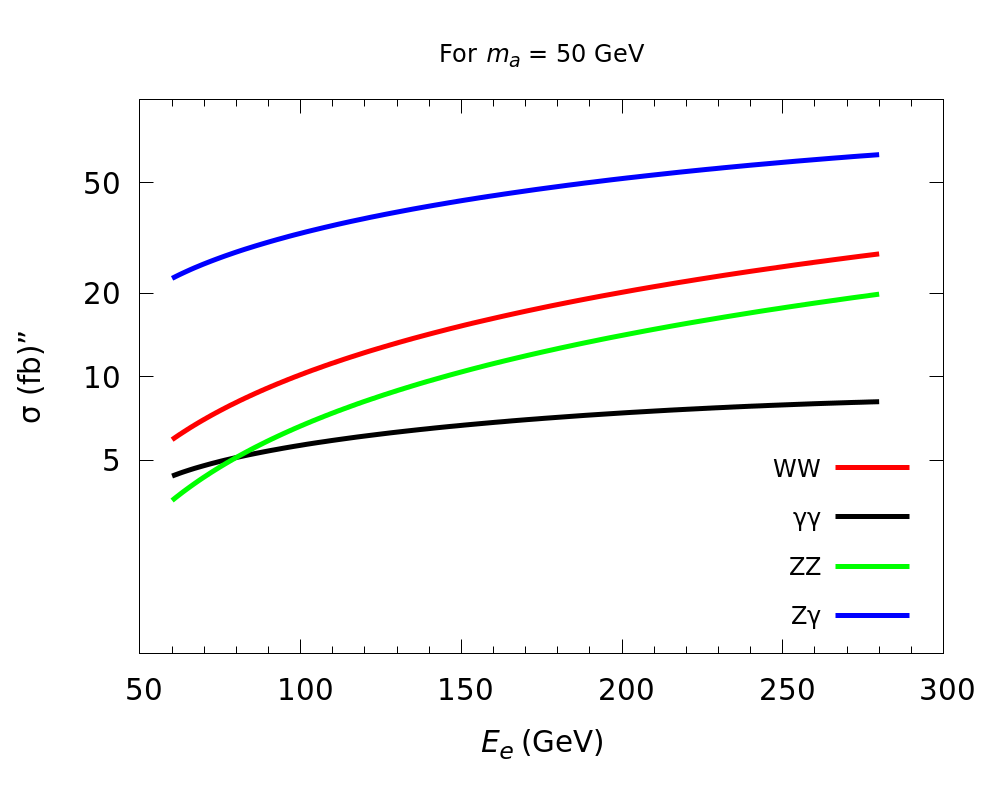}}
\caption{The production cross section of ALP production in {\tt{CC}} and {\tt{NC}} where $a\to \gamma\gamma$ as a function of (a) ALP-mass $m_a$, and (b) electron-energy $E_e$ for fixed energy of proton $E_p = 7$~TeV. Note that the ${\tt Z}\gamma$-channel can not be separated from $\gamma\gamma$-channel for Case (I) and hence here $Z\gamma$ represents total cross section considering the channels shown in~\autoref{fig:yy},~\autoref{fig:Zy} and their interference. However, in Case (II), contributions from the ${\tt ZZ}$-channel are also included. Solid (dashed) lines represent Case (I) (Case (II)). \label{fig:2}}
\end{figure*}

To explore the goals of this study, we first build a model file for the interactions defined in~\autoref{eq:2} using the package {\tt{FeynRules}}~\cite{Alloul:2013bka}. 
For the generation of events, we use the Monte Carlo event generator package \MGvATNLO~\cite{Alwall:2011uj}.  Further showering, fragmentation and hadronization are done with a customized {\tt{Pythia-PGS}}~\cite{Sjostrand:2006za}, and the detector level simulation performed with reasonably chosen parameters using {\tt{Delphes}} \cite{deFavereau:2013fsa} and jets were clustered using {\tt{FastJet}}~\cite{Cacciari:2011ma} with the anti-$k_T$ algorithm~\cite{Cacciari:2008gp} using the distance parameter, $R = 0.4$ as explained in ref.~\cite{Kumar:2015kca}. The factorization and normalization scales are set to be dynamic scales for both signal and potential backgrounds. For this study, $e^-$ polarization is assumed to be $-80$\%. The initial requirements on transverse momentum ($p_T$) and rapidity ($\eta$) of jets, leptons and photons are nominal: $p_T^{j,  e^-, \gamma} > 10$~GeV, $|\eta_{j, e^-,  \gamma}| < 5$ and no cuts on missing energy.

With these setups the estimated cross-section of ALP production through (a) {\tt {CC}} process: $e^-p \to \nu_e a j$, and (b) {\tt {NC}} process: $e^-p \to e^- a j$ 
with further decay of $a \to \gamma\gamma$ in the mass range of $5 \le m_a \le 300$~GeV is shown in~\autoref{xsec-ma} for a benchmark electron's energy $E_e = 60$~GeV and proton energy $E_p = 7$~TeV at LHeC. Also for a fixed $m_a = 50$~GeV, cross-section as a function of $60 \le E_e \le 300$~GeV is shown in~\autoref{xsec-ee}. Note that for the $WW$, $\gamma\gamma$ and $ZZ$-fusion the corresponding coupling value is taken as $g_{kk}=1$ ($k = W, \gamma, Z$) keeping others $0$, while for $Z\gamma$-fusion $g_{Z\gamma} = g_{\gamma\gamma} = 1$ keeping $g_{WW} = 0 = g_{ZZ}$ as stated in Case (I) (solid lines). For Case (II) (dashed lines), cross sections for ${\tt WW}$, $\gamma\gamma$ and ${\tt ZZ}$ channels keep decreasing for $m_a > m_Z$, while for $Z\gamma$ its overall higher than Case (I) due to ${\tt ZZ}$ contributions. Since $m_a < m_Z$ in the case of~\autoref{xsec-ee}, Case (II) has no effect.

In next~\autoref{sec:results}, we will focus on background generation and analysis procedures to estimate the bounds on the couplings $g_{ij}$ using the methods described in~\autoref{sec:model}. We will construct observables that are sensitive to the presence of these couplings and use them to establish the limits. 

\section{Analysis, observable and results}
\label{sec:results}
\begin{figure}[t]
\centering
%\subfloat[\label{}]
{\includegraphics[width=0.50\textwidth]{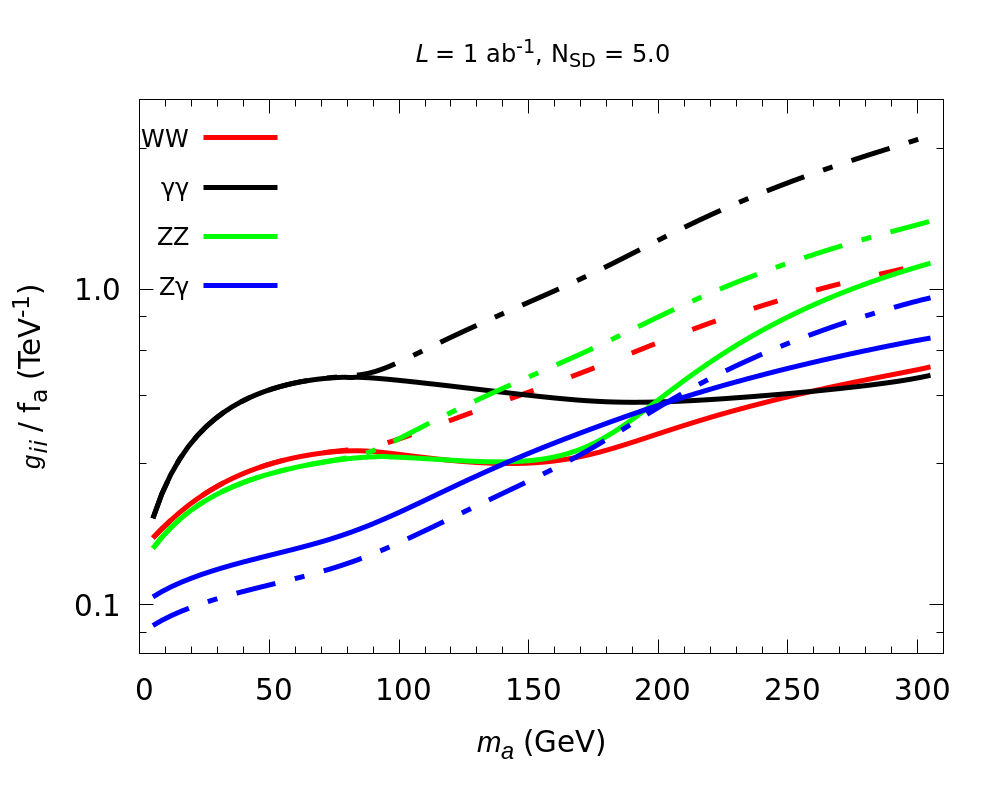}}
\caption{The projected 5$\sigma$ sensitivities for $g_{ij}/f_a$ by using the formula in~\autoref{eq:8}, on optimised events (see text for details). Solid (dashed) lines represent Case (I) (Case (II)).}
\label{fig:3}
\end{figure}
\begin{figure*}[t]
\centering
\subfloat[\label{d-ww}]{\includegraphics[width=.50\textwidth]{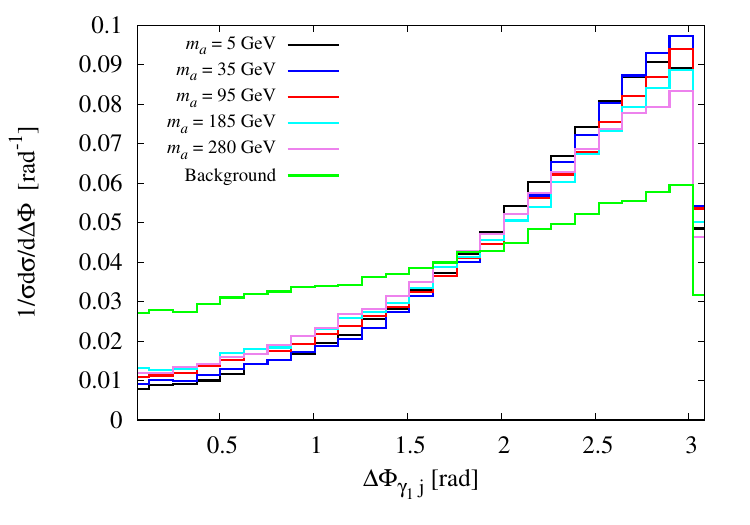}}
\subfloat[\label{d-yy}]{\includegraphics[width=.50\textwidth]{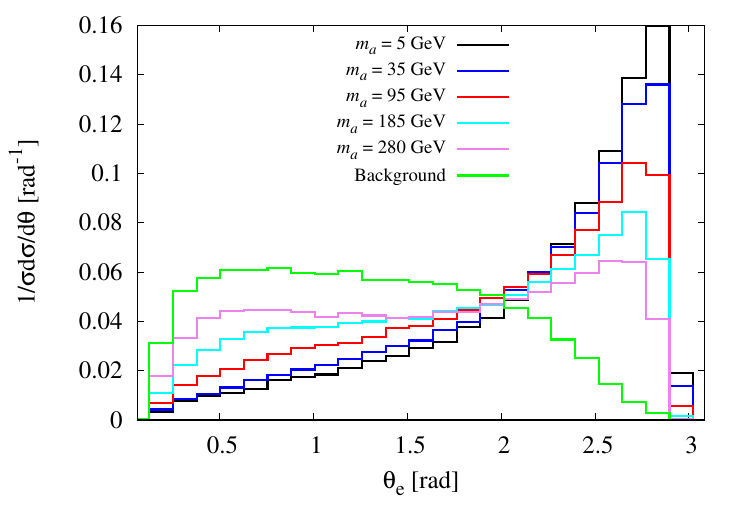}} \\
\subfloat[\label{d-zz}]{\includegraphics[width=.50\textwidth]{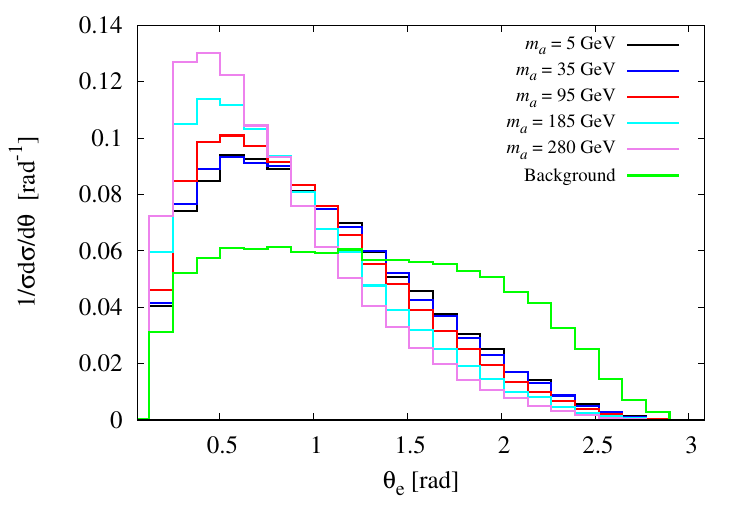}}
\subfloat[\label{d-zy}]{\includegraphics[width=.50\textwidth]{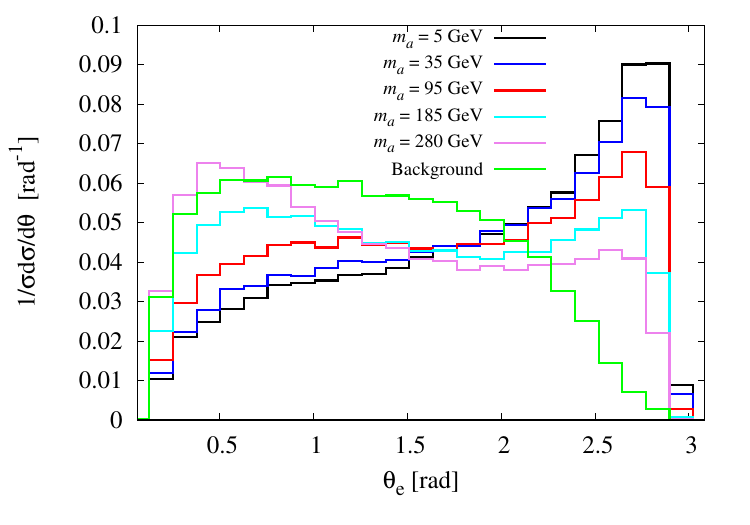}}
\caption{Representative normalized differential distributions for Case (I) with 80\% left polarized $e^-$ beam for (a) {\tt{WW}}: $\Delta \Phi_{\gamma_1 j}$ - the azimuthal angle between the two planes of the final state leading-$p_T$ photon ($\gamma_1$) and forward jet with respect to the beam direction, and the scattered angle $\theta_e$ with respect to beam direction for the final state tagged $e^-$ for (b) $\gamma \gamma$, (c) {\tt{ZZ}}, (d) $Z\gamma$ channels, where five benchmarks $m_a$ signal events are shown with respect to the dominant background using $E_p = 7$~TeV and $E_e = 60$~GeV with selection cuts explained in texts. %In all cases $g_{ij} = 1$ and ${\cal B}_{a\to\gamma\gamma} = 1$.
}
\label{fig:4}
\end{figure*}
\begin{figure*}[t]
\centering
\subfloat[\label{fig:5a}]{\includegraphics[width=.5\textwidth]{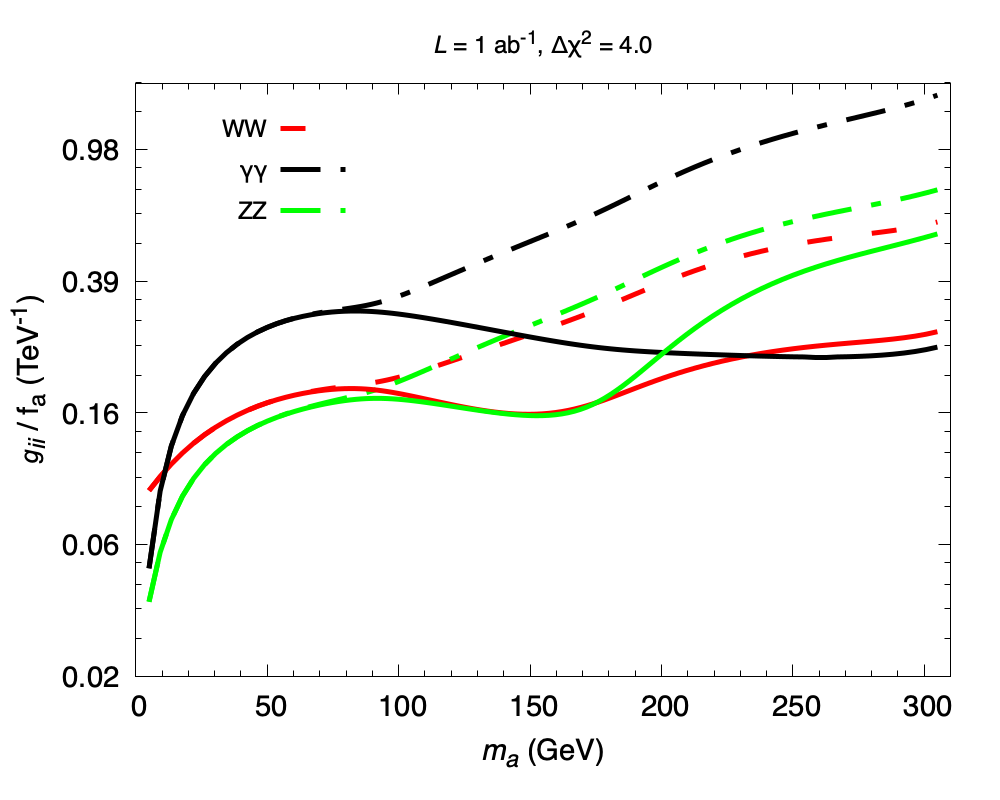}}
\subfloat[\label{fig:5b}]{\includegraphics[width=.5\textwidth]{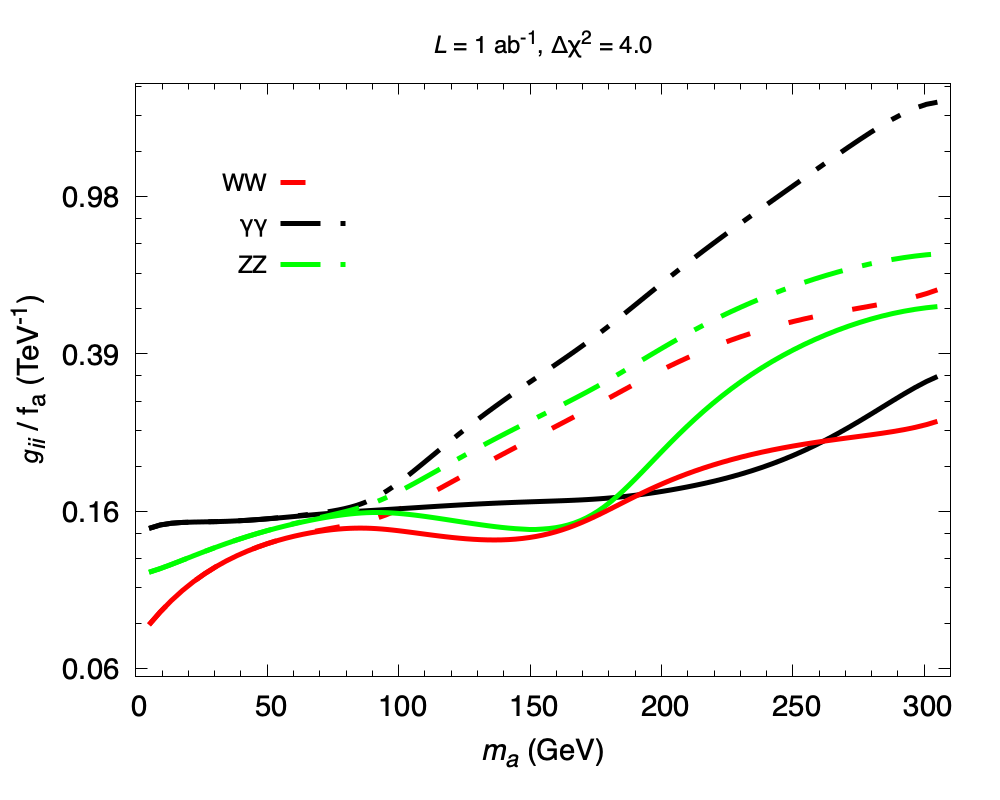}}
\caption{The 95\% C.L. contours are shown in the $g_{ii}/f_a - m_a$ plane with the observable based on $\chi^2$-analysis for (a) one-bin and (b) multiple-bin with integrated luminosity of $L = 1$~ab$^{-1}$. Solid (dashed) lines represent Case (I) (Case (II)).
\label{fig:5}}
\end{figure*}
\begin{figure*}
\centering
\subfloat[\label{fig:6a}]{\includegraphics[width=.5\linewidth]{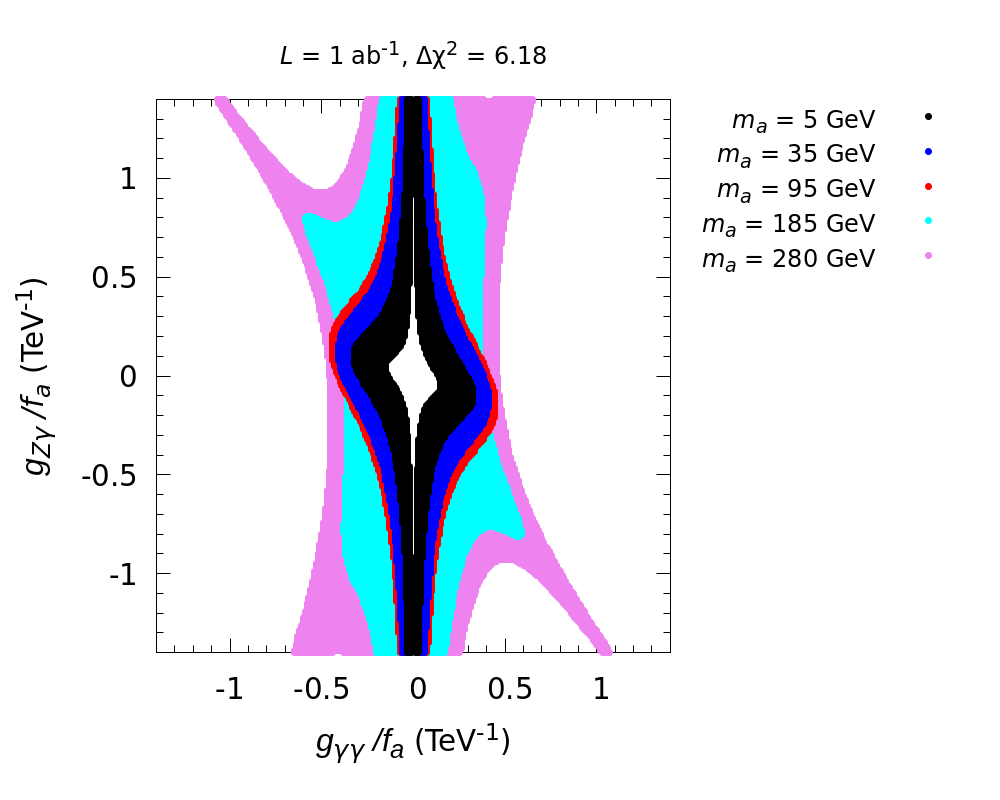}}
\subfloat[\label{fig:6b}]{\includegraphics[width=.5\linewidth]{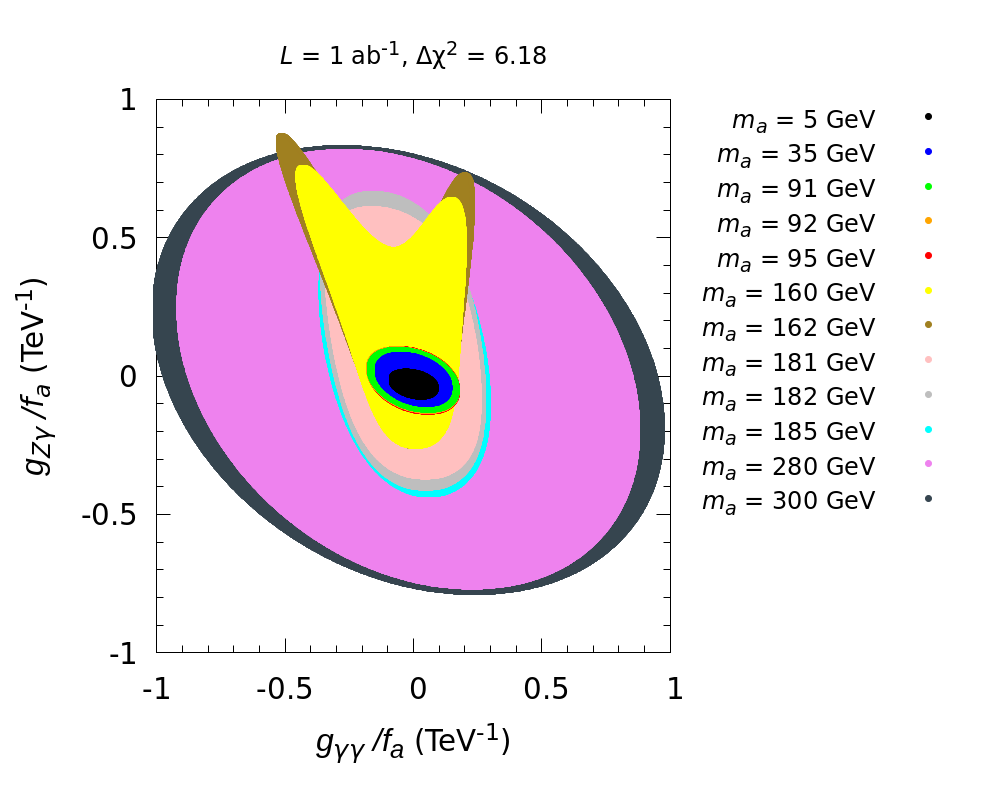}}
\caption{The 95\% C.L. contours are shown in the $g_{Z\gamma}/f_a - g_{\gamma\gamma}/f_a$ plane for selective $m_a$ in (a) Case (I) and (b) Case (II) considering $g_{ZZ} = 0.1$;  with the observable based on multiple-bins $\chi^2$-analysis as explained in text with integrated luminosity of $L = 1$~ab$^{-1}$.} \label{fig:6}
\end{figure*}

To generate backgrounds, we adopt similar setups as described earlier. This includes specifying the center-of-mass energy, beam polarization, and luminosity, as well as considering the relevant physics processes with ``di-photon + jets'' final state in {\tt{CC}}, {\tt{NC}} and photo-production modes and their corresponding cross sections. For $E_e = 60$~GeV and $E_p = 7$~TeV, the estimated total cross-section of background (signal) is approximately less than 6~fb (shown in~\autoref{xsec-ma} as a function of $m_a$). To optimise the signal events over the leading backgrounds additional cuts on leading and sub-leading jets, photons and leptons are applied depending on channels in this study:
\begin{itemize}
    \item for all channels: $p_T^{j,  e^-, \gamma} > 20$~GeV,
    \item {\tt{WW}}: $0 <\eta_\gamma < 3$, $0 < \eta_j< 4$,
    \item {\tt{$\gamma\gamma$}}: $-2 < \eta_\gamma < 3$, $-2 < \eta_j < 5$, $-2.5 < \eta_{e^-} < 2$, 
    \item {\tt{ZZ}}: $0 < \eta_\gamma< 3$, $0 <  \eta_j < 5$, $0 < \eta_{e^-}< 5$, and
    \item {\tt{Z$\gamma$}}: $0 <  \eta_\gamma < 3$, $0 < \eta_j < 5$, $-2.5 < \eta_{e^-} < 1$,
    \end{itemize}
in addition with invariant di-photon mass, $m_{\gamma\gamma}$, cuts as a function of corresponding signal of $m_a$ in the window of $\sim \pm 5$~GeV. This cut significantly reduces the backgrounds in comparison to signal events. By having these optimized events we then estimate the significance and evaluate the projected sensitivities by using the formula in~\autoref{eq:8}. In~\autoref{fig:3} we show the discovery limit on the coupling $g_{ij}/f_a$ as a function of $m_a$ by fixing $N_{\rm SD} = 5$. These limits are direct reflection of cross-section (and branching ratio) dependence on $m_a$ as shown in~\autoref{fig:2} (\autoref{fig:br}).

Though we need to find a mechanism through which these limits must improve, and for that we studied various possible observables by considering the differential distributions and combinations of tagged final state $e^-$, photons, and jets for the signal as well backgrounds. In~\autoref{fig:4} we show most sensitive normalized differential distributions (for Case (I) only as representative), where for {\tt{WW}} channel, $\Delta\Phi_{\gamma_1 j}$, the azimuthal angle between the two planes of the final state leading-$p_T$ photon and forward jet with respect to the beam direction is shown in~\autoref{d-ww}. However, the scattered angle $\theta_e$ with respect to beam direction for the final state tagged $e^-$ is most sensitive for $\gamma\gamma$, {\tt{ZZ}} and $Z\gamma$ channels shown in~\autoref{d-yy}, \autoref{d-zz} and \autoref{d-zy}, respectively. It is interesting to note that the signal events in case of $\gamma\gamma$ ({\tt{ZZ}})-channel lies towards higher (lower) $\theta_e$ due to pure QED ($V$-$A$) structure of photon-fermion ($Z$-fermion) couplings, though its mixed in case of $Z\gamma$ channel. And the shape of backgrounds are due to the selection of different $\eta$-regions. So furthermore we perform a $\chi^2$-analysis at both cross section (one-bin) and differential distribution (multiple-bin) levels \footnote{A one-bin $\chi^2$ analysis refers to the calculation of $\chi^2$ for the total cross section, where the entire distribution is considered as a single bin. This approach combines all the observed and expected values across all bins and calculates the $\chi^2$ based on the overall distribution. In a multiple-bin $\chi^2$ analysis, the observed data are divided into different bins based on the values of the kinematic observable. The expected theoretical distribution (SM background) is also divided into the corresponding bins. The $\chi^2$ value is then calculated for each bin by comparing the observed and expected values, taking into account the uncertainties or errors in the observed data. The individual $\chi^2$ values for each bin are typically summed to obtain the total $\chi^2$ value for the analysis. Therefore a multiple-bin $\chi^2$ analysis captures the differential information present in each bin separately, providing more detailed insights into the distribution across different kinematic regions. In contrast, a one-bin $\chi^2$ analysis provides an overall measure of the goodness-of-fit but does not account for the variations or discrepancies within individual bins.} and apply~\autoref{eq:9} on these observable to estimate the sensitivities of $g_{ij}/f_a$ as a function of $m_a$. 

%for $\chi^2 \le 3.09$.
Note that the general structure of the $\sigma(g_{ij})$ is given as 
\begin{align}
    &\sigma(g_{ii}) = g_{ii}^2 \sigma_{ii} \times {\cal{B}}_{a\to\gamma\gamma};  \notag\\
    &\sigma (g_{Z\gamma}) = \Big(g_{\gamma\gamma}^2 \sigma_{\gamma\gamma} + g_{ZZ}^2 \sigma_{ZZ} + g_{Z\gamma}^2 \sigma_{{\tt{Z}}\gamma} + g_{\gamma\gamma} g_{ZZ} \sigma_{\rm inf.}^{(1)}  \notag \\
    &\qquad\qquad + g_{ZZ}g_{Z\gamma} \sigma_{\rm inf.}^{(2)} + g_{\gamma\gamma}g_{Z\gamma} \sigma_{\rm inf.}^{(3)}\Big) \times {\cal{B}}_{a\to\gamma\gamma}. 
\label{Xsecform}
\end{align}
For Case (I), $g_{ZZ} = 0$ and \autoref{Xsecform} provides the justification for the Mexican-hat shape of the $\chi^2$ distribution, where the minimum value is $\chi^2_{\rm min.} = 0$. In the case of a one-parameter analysis of $g_{ii}/f_a$ {\it{vs}} $m_a$, we set $\chi^2 \equiv \Delta\chi^2 = 4.0 $ to correspond to the 95\% confidence level (C.L.). In~\autoref{fig:5a} and~\autoref{fig:5b}, we present the sensitivities of $g_{ii}/f_a$ {\it{vs}} $m_a$ using the one-bin and multiple-bin $\chi^2$ analyses, respectively. From~\autoref{fig:5a}, it is evident that the limits on $g_{ii}/f_a$ perform significantly better compared to those obtained using~\autoref{eq:8} (as shown in~\autoref{fig:3}). However, the multiple-bin analysis on differential distributions, as shown in~\autoref{fig:5b}, outperforms the one-bin analysis (\autoref{fig:5a}) specifically for the {\tt{WW}} and $\gamma\gamma$ channels. This indicates that considering multiple bins in the analysis provides improved sensitivity in constraining the values of $g_{ii}/f_a$ {\it{vs}} $m_a$ for these specific channels. The results presented in~\autoref{fig:5a} and~\autoref{fig:5b} demonstrate the enhanced performance of the multiple-bin analysis approach, emphasizing its superiority in capturing the sensitivity to $g_{ii}/f_a$ {\it{vs}} $m_a$ compared to the one-bin analysis, particularly for the {\tt{WW}} and $\gamma\gamma$ channels.

Since the multiple-bin analysis performs better in these three scenario, in~\autoref{fig:6a}, the limits for the $Z\gamma$ channel in the $g_{\gamma\gamma}/f_a - g_{Z\gamma}/f_a$ plane {\it{vs}} $m_a$ are shown for five selected values of $m_a$ for $\chi^2 \equiv \Delta \chi^2 = 6.18$ (as of two-parameter analysis for 95\% C.L.) using this approach only. It is observed that the shape of the limits is asymmetric with respect to $g_{\gamma\gamma} = g_{Z\gamma} \approx 0$, where the limits also blow up. This asymmetry can be attributed to the presence of negative and positive interference effects. According to~\autoref{Xsecform}, the region around $g_{\gamma\gamma} = g_{Z\gamma} \approx 0$ can be understood. In this region, all values of $g_{Z\gamma}$ can satisfy the $\chi^2$ criterion below the 2$\sigma$ standard deviation when $g_{\gamma\gamma}$ tends to zero. However, we have excluded the region near $g_{\gamma\gamma} = 0$ in order to fulfill the minimum requirement of an ALP signal for the study. The observed spikes in the contour is due to the negative contribution from interference, which leads to infinite values for both couplings. The presence of four spikes can be attributed to the even powers of the couplings in the cross-section~\autoref{Xsecform}. When the value of $g_{\gamma\gamma}$ is non-zero, these spikes disappear, and the contour takes on a circular shape due to the negligible contribution from interference.

In~\autoref{fig:6b}, limits on the $g_{\gamma\gamma}/f_a - g_{Z\gamma}/f_a$ plane {\it{vs}} $m_a$ are presented for Case (II). To achieve a 95\% C.L. corresponding to $\Delta \chi^2 = 6.18$, we establish a benchmark point by setting $g_{ZZ} = 0.1$. This choice is made because, for $g_{ZZ} = 1$, corresponding values of $\Delta \chi^2$ exceed 6.18.  Significant deviations at the mass points for $m_a$ corresponding to the $Z\gamma$, $W^+W^-$, and $ZZ$ resonances are readily apparent.
\begin{figure*}[t]
\centering
\subfloat[\label{fig:7a}]{\includegraphics[width=.50\textwidth]{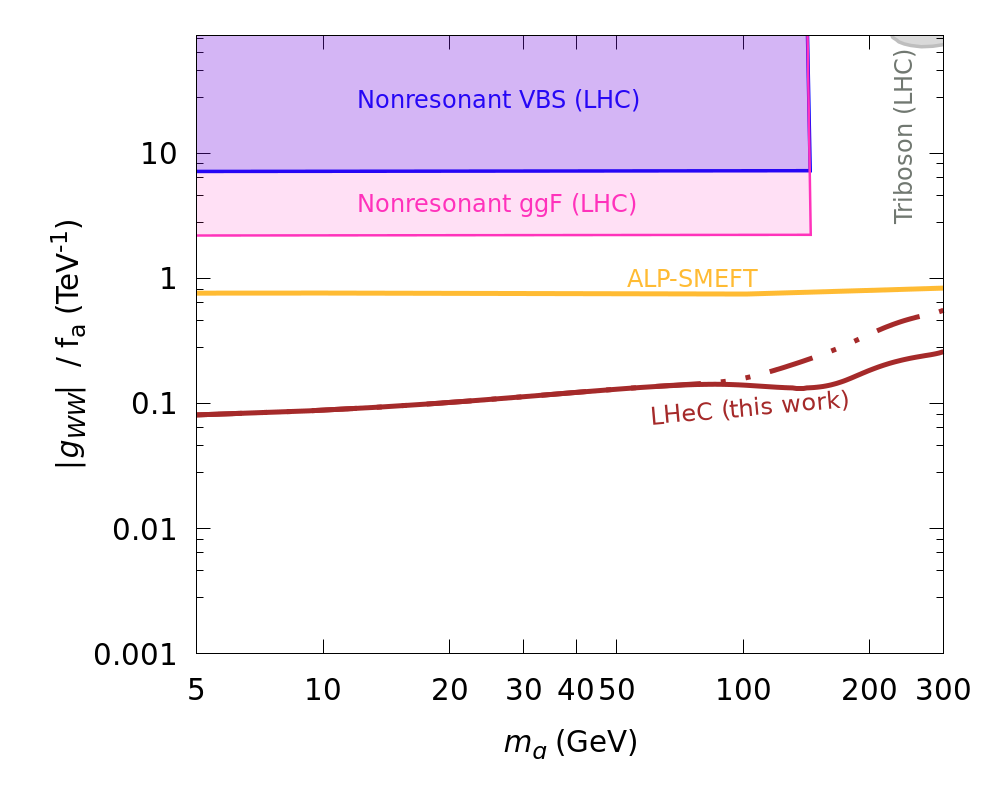}}
\subfloat[\label{fig:7b}]{\includegraphics[width=.50\textwidth]{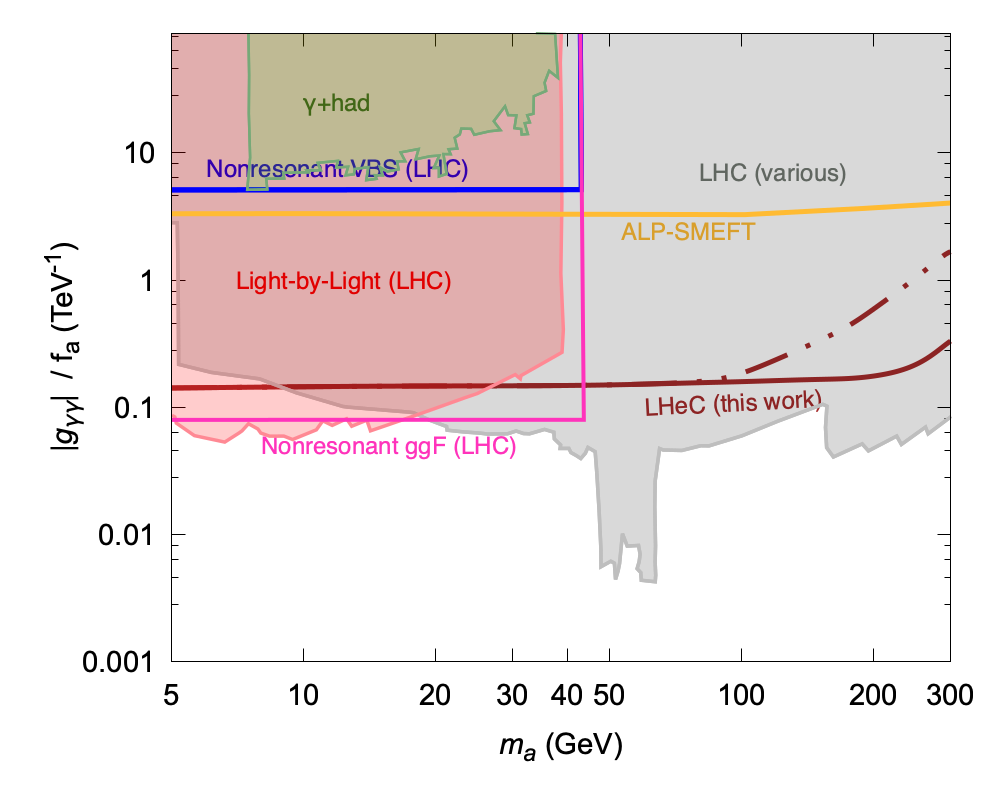}}\\
\subfloat[\label{fig:7c}]{\includegraphics[width=.50\textwidth]{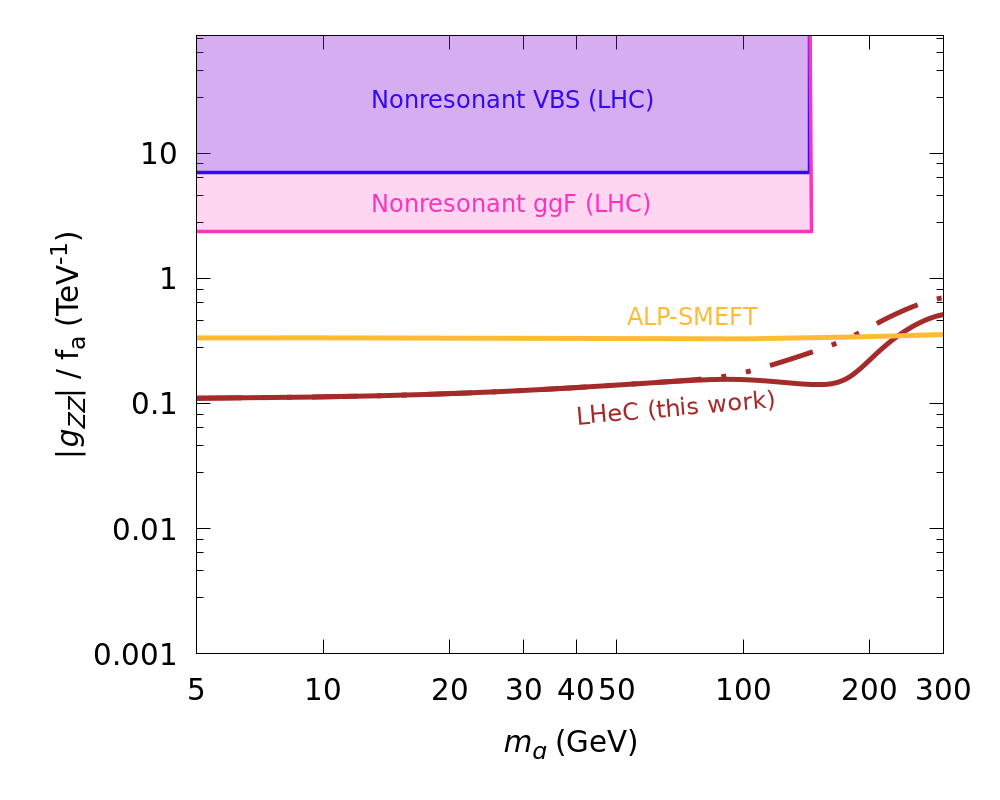}}
\subfloat[\label{fig:7d}]{\includegraphics[width=.50\textwidth]{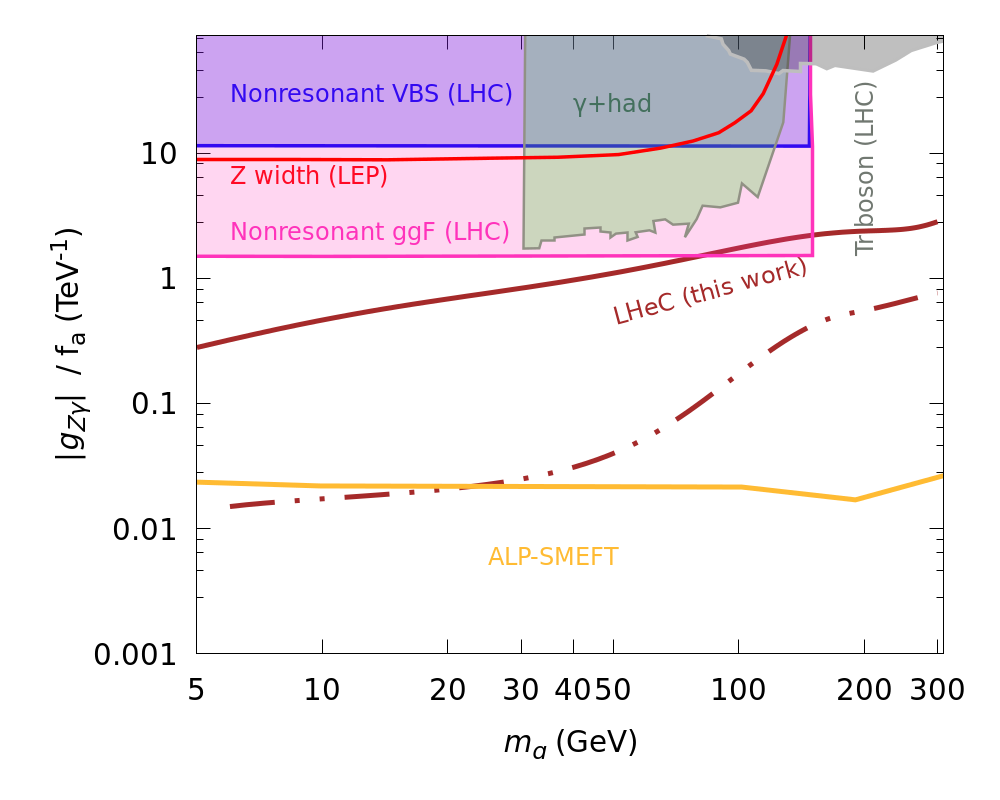}}
\caption{The 95\% C.L. contours in the $\left|g_{ij}\right|/f_a - m_a$ plane are shown for the limits obtained from multiple-bin $\chi^2$-analysis with an integrated luminosity of $L = 1$~ab$^{-1}$. Corresponding available limits from LHC, $\gamma+$hadron, $Z$ width at LEP, and ALP-SMEFT are also shown for comparison (see text for details). Our constraints given in~\autoref{fig:6a} (\autoref{fig:6b}) for $g_{Z\gamma}$ are correlated with $g_{\gamma \gamma}$ (and $g_{ZZ}$) due to the interference. For standalone comparison of constraints on $g_{Z\gamma}$ with previous studies, we keep $g_{\gamma \gamma}=0.1$ (and $g_{ZZ} = 0.1$); where brown solid (dashed) line represents Case (I) (Case (II)).}\label{fig:7}
\end{figure*}

\section{Comparison of $g_{ij} (m_a)$ to existing bounds}
\label{comp}
In~\autoref{fig:7}, a comparison of coupling limits is presented in the $\left|g_{ij}\right|/f_a - m_a$ plane at the 95\% confidence level (C.L.), along with constraints from various experiments and theory predictions. It is important to note that a given measurement can depend on multiple ALP couplings. Representing the corresponding bound in the 2D ($\left|g_{ij}\right|/f_a$, $m_a$) plane requires making theoretical assumptions, which can vary significantly from constraint to constraint. These differences should be considered for a proper comparison. In~\autoref{fig:7}, the bounds derived in the present work (shown as the brown line) represent the 95\% C.L. limits. They are derived assuming full decay of the ALP to di-photons. In order to compare the limits on $Z\gamma$-channel, given in~\autoref{fig:6a} (\autoref{fig:6b}) as a correlation between $g_{Z \gamma}$ and $g_{\gamma \gamma}$ (and $g_{ZZ}$) due to the interference, we show a standalone comparison of constraints on $g_{Z\gamma}$ with previous studies keeping $g_{\gamma \gamma} = 0.1$ (and $g_{ZZ} = 0.1$) for Case (I) (Case (II)).

The limits on $g_{\gamma \gamma}$ and $g_{Z\gamma}$ at higher ALP masses are obtained from collider studies, where the ALP decays resonantly either to hadrons or to photon pairs. The relevant process are from $e^+ e^- \rightarrow \gamma$ + hadrons, studied by the L3 experiment~\cite{L3:1992kcg} and the leading bounds from photon pair production at the Large Hadron Collider (LHC) in proton-proton collisions~\cite{Jaeckel:2012yz, Mariotti:2017vtv} (labeled as ``LHC'' for measurements from ATLAS and CMS), as well as from light-by-light scattering $\gamma \gamma \to \gamma \gamma$ measured in lead-lead (Pb-Pb) collisions~\cite{CMS:2018erd, ATLAS:2020hii} (labeled as ``Light-by-light (LHC)''). The measurement of the total $Z$ decay width at LEP provides constraints up to $m_a \lesssim m_Z$~\cite{Brivio:2017ije, Craig:2018kne}.

For ALP masses above 100~GeV, the dominant bounds come from resonant triboson searches~\cite{Craig:2018kne}. Additionally, nonresonant searches in diboson production via gluon fusion at the LHC (labeled as ``Nonresonant ggF'') provide constraints on all four ALP interactions. Each nonresonant bound is extracted from a specific process $gg \rightarrow a^* \rightarrow V_1 V_2$ ($V = \gamma, Z, W^\pm$). The constraint on $g_{\gamma \gamma}$ is derived in ref.~\cite{Gavela:2019cmq}, those on $g_{WW}$ and $g_{\gamma Z}$ in ref.~\cite{Carra:2021ycg}, and the constraint on $g_{ZZ}$ in ref.~\cite{CMS:2021xor}.

The bound obtained from the $Z$ width measurement at LEP does not require additional assumptions. The bounds from nonresonant ggF, which include nonresonant $gg \to a^* \rightarrow V_1 V_2$ processes, scale with the inverse of the axion-gluon coupling ($g_{gg}$) and are completely lifted when the ALP coupling to gluons $C_{GG} \rightarrow 0$. In the figure, they are normalized to $g_{gg} = 1$ (for details see ref.~\cite{Bonilla:2022pxu}).

Bounds labeled as ``$\gamma$ + had'' and LHC (various) assume gluon dominance, i.e., $g_{gg} \gg g_{V_1 V_2}$, and in this limit, they are largely independent of $C_{GG}$ (see ref.~\cite{Alonso-Alvarez:2018irt}). Among these, bounds on $g_{\gamma \gamma}$ labeled as ``LHC" additionally assume negligible branching fractions to fermions and heavy electroweak bosons in the mass region where they are kinematically allowed. The limit from light-by-light scattering, shown in red, assumes ${\cal B}_{a \rightarrow \gamma \gamma}$ = 1, which corresponds to vanishing couplings to gluons and light fermions. The triboson constraints on $g_{WW}$ and $g_{Z \gamma}$ make use of the photophobic ALP scenario~\cite{Craig:2018kne}.

A recent study has utilized ALP-SMEFT interference to establish limits on various ALP couplings~\cite{Biekotter:2023mpd}. These bounds exhibit particular effectiveness in the ALP mass range where constraints from flavor and astrophysics searches tend to weaken. We have applied these bounds to our specific cases using \autoref{eq:3} and compared the resulting limits in~\autoref{fig:7}.

Limits on the effective ALP-photon coupling have been derived from exotic Higgs and $Z$ decay searches at the LHC in~\cite{Bauer:2017ris,Bauer:2018uxu}, and the probed parameter space is found to be in agreement with the results presented in this work.

Overall the limits found in this work performs better sensitivity for all three ALP couplings, namely, $g_{WW}$, $g_{ZZ}$ and $g_{Z\gamma}$ comparing to available studies in different collider scenario, whereby, the limits on $g_{\gamma\gamma}$ are competitive with respect to few cases. In ALP-SMEFT bounds, the performance of $g_{Z\gamma}$ is relatively poor.  

\section{Summary and discussions}
\label{sec:sum}

In this article, we investigated the potential for the production of relatively high-mass Axion-Like Particles (ALPs) in an electron-proton ($e^-p$) environment. Specifically, we focused on the proposed energy of the Large Hadron-electron Collider (LHeC) with a center-of-mass energy of  $\sqrt{s} \approx 1.3$~TeV and an integrated luminosity of $L = 1$~ab$^{-1}$. Although exploring high masses beyond 300~GeV is less likely due to the limited cross section achievable with the available energy and luminosity, we examined the limits on coupling measurements as prediction for such masses based on our analysis procedure. These limits serve as approximate predictions that can be investigated further if the electron energy ($E_e$) is increased to higher values.

In~\autoref{fig:7} we provide a comprehensive overview of the coupling limits in the $\left|g_{ij}\right|/f_a-m_a$ plane, taking into account various experimental and theoretical constraints, and highlights the strengths and limitations of each measurement in constraining ALP couplings. To analyze and capture the differential information in the distribution of kinematic observables, a multiple-bin $\chi^2$ analysis is preferable in contrast to one-bin, where we observed the limits performance are better. Also the limits on $g_{WW}$, $g_{ZZ}$ and $g_{Z\gamma}$ comparing to available studies in different collider scenario are better at LHeC for considered range of $m_a$, whereby, the limits on $g_{\gamma\gamma}$ are competitive with respect to few scenario.

By studying the possibilities of ALP production in the $e^-p$ environment at the LHeC, we contribute to the understanding of ALP physics and provide insights into the potential for probing relatively higher masses and coupling strengths in future experiments. While numerous studies on probing ALPs have been conducted, this article stands among the first to explore the potential of a proposed future $e^-p$ collider, specifically at the suggested energy levels of the LHeC.

%\acknowledgement 
\begin{acknowledgements}
PS would like to acknowledge the School of Physics at the University of the Witwatersrand, where the majority of this work was conducted during his visit. Their support and resources were invaluable in carrying out this research. AG thanks SERB , G.O.I. Under CRG/2018/004889. 
\end{acknowledgements}
{\bf{Data Availability Statement}} This manuscript has no associated dat or the data will not be deposited. [Authors’ comment: For this work authors used publicly available data where necessary, and are referenced
properly.]

\bibliographystyle{unsrt}

\end{document}